\documentclass[aps,prmaterials,superscriptaddress,numerical,reprint,showkeys,floatfix]{revtex4-1}

\makeatletter 

\usepackage{verbatim,amsmath,amssymb,tabularx,multirow,graphicx,float,txfonts}
\usepackage[caption=false]{subfig}
\usepackage[utf8]{inputenc}
\usepackage{color}
\usepackage{natmove}
\usepackage{makecell}
\graphicspath{{./figures/}{./}}

\begin{document}
\title{Stability of charged sulfur vacancies in 2D and bulk MoS$_2$ from plane-wave density functional theory with electrostatic corrections}
\author{Anne Marie Z. Tan}
\email{annemarietan@ufl.edu}
\affiliation{Department of Materials Science and Engineering, University of Florida, Gainesville, FL 32611, USA}
\affiliation{Quantum Theory Project, University of Florida, Gainesville, FL 32611, USA}
\author{Christoph Freysoldt}
\affiliation{Max-Planck-Institut f{\"u}r Eisenforschung GmbH, Max-Planck-Stra{\ss}e 1, 40227 D{\"u}sseldorf, Germany}
\author{Richard G. Hennig}
\email{rhennig@ufl.edu}
\affiliation{Department of Materials Science and Engineering, University of Florida, Gainesville, FL 32611, USA}
\affiliation{Quantum Theory Project, University of Florida, Gainesville, FL 32611, USA}
\date{\today}

\begin{abstract}

Two-dimensional (2D) semiconducting transition metal dichalcogenides such as MoS$_2$ have attracted extensive research interests for potential applications in optoelectronics, spintronics, photovoltaics, and catalysis. To harness the potential of these materials for electronic devices requires a better understanding of how defects control the carrier concentration, character, and mobility. 
Utilizing a correction scheme developed by Freysoldt and Neugebauer to ensure the appropriate electrostatic boundary conditions for charged defects in 2D materials, we perform density functional theory calculations to compute formation energies and charge transition levels associated with sulfur vacancies in monolayer and layered bulk MoS$_2$. We investigate the convergence of these defect properties with respect to vacuum spacing, in-plane supercell dimensions, and different levels of theory.
We also analyze the electronic structures of the defects in different charge states to gain insights into the effect of defects on bonding and magnetism. We predict that both vacancy structures undergo a Jahn-Teller distortion, which helps stabilize the sulfur vacancy in the $-1$ charged state.
\end{abstract}

\maketitle

\section{Introduction}

Two-dimensional (2D) semiconductor materials, such as transition metal dichalcogenides (TMDCs), monochalcogenides, group III-V compounds, and phosphorene, have attracted extensive research interests for potential applications in optoelectronics, spintronics, photovoltaics, and catalysis \cite{mak2010, radisavljevic2011, geim2013, butler2013, Zhuang2013, Zhuang2013-SnS2, jariwala2014,  liu2014, li2014, xia2014, wu2015, ye2015, cao2015, Singh2015, albalushi2016, Zhuang2016, Ashton2017, Paul2017}. One of the most commonly used and promising 2D semiconductor materials is the prototypical TMDC material, molybdenum disulfide (MoS$_2$), which has demonstrated interesting electronic, optical, and mechanical properties, making it a promising candidate for optoelectronic and catalytic applications \cite{mak2010,radisavljevic2011,lopez2013, Zhuang2017, li2019}. Monolayer MoS$_2$ can be directly grown on substrates using chemical vapor deposition \cite{lee2012} or exfoliated from its layered bulk counterpart via micromechanical \cite{novoselov2005, radisavljevic2011} or liquid-phase \cite{joensen1986, coleman2011} exfoliation techniques. In this work, we focus on the semiconducting trigonal prismatic 1H-MoS$_2$ phase and its layered bulk counterpart, 2H-MoS$_2$.  

Just as in bulk semiconductors, 2D semiconductors contain both intrinsic defects, {\it e.g.}, vacancies and antisites, as well as extrinsic defects, {\it e.g.}, substitutional and interstitial dopants and impurities. These defects are often charged and can also interact to form pairs or complexes. The lower dimensionality of 2D materials reduces the electronic screening, and hence point defects are expected to have an even stronger impact on the electronic properties of these systems compared to in bulk semiconductors. Understanding the effect of defects, dopants, and impurities on the electronic properties is crucial for the selection of materials and the choice of suitable synthesis and processing conditions. Accurate determination of defect formation energies and charge transition levels (CTLs) enables us to predict their effect on the electronic properties and how they respond to changes in synthesis and processing, allowing for some control over the defect concentrations, and hence to tailor the carrier concentration, character, and mobility in 2D materials \cite{peng2013, lin2016, zhao2016, yang2016}. 

Unfortunately, experimental data of defect concentrations are scarce due to the difficulty of measuring defects in low-dimensional systems and establishing and maintaining thermodynamic equilibrium in these systems.
Exciton emission peaks associated with defect states within the bandgap have been observed in the photoluminescence spectra of MoS$_2$~\cite{tongay2013,chow2015,zhang2018}; however, the type and nature of the defects responsible for such peaks are not directly known and have to be inferred by comparison against computational predictions of defect levels. This highlights the importance of accurate computational studies of defect levels and formation energies to complement experimental observations to better understand the effect of defects, dopants, and impurities on the electronic properties of emerging 2D semiconductor materials. 

Density functional theory (DFT) calculations of point defects in solids is a mature field with a proven record of experimentally validated predictions \cite{freysoldt2014}. Similar approaches may be applied to point defects in 2D materials as well; however, additional care must be taken to ensure the appropriate electrostatic boundary conditions for charged defects in 2D materials when applying DFT methods utilizing plane-wave basis sets and periodic boundary conditions. Several previous computational studies on defects and dopants in MoS$_2$ considered neutral defects \cite{haldar2015,dolui2013}, or employed the Lany-Zunger (modified Markov-Payne) correction \cite{lany2008,liu2013,lu2018}, or a uniform scaling scheme \cite{komsa2012prb,komsa2015,noh2014} to treat the charged defects.

In this work, we use the 2D charge correction scheme developed recently by Freysoldt and Neugebauer \cite{freysoldt2018} -- which has an advantage of not requiring any extrapolation or prior knowledge of scaling laws -- to compute the formation energies and CTLs associated with the single S vacancy in both monolayer and layered bulk MoS$_2$. The S vacancy has been observed directly by high-resolution transmission electron microscopy \cite{komsa2012prl,zhou2013,hong2015} and scanning tunneling microscopy \cite{vancso2016} and has been predicted to have one of the lowest formation energies compared to other intrinsic defects \cite{noh2014, komsa2015, haldar2015}. Since the S vacancy is one of the more well-studied defects in MoS$_2$, we use it here to validate and benchmark this new charge correction scheme for 2D monolayers as well as the SCAN+rVV10 functional \cite{SCANrVV10-peng2016}.

The paper is organized as follows. Sec.~\ref{comp-details} lays out the DFT computational details and benchmarks the calculated structural and electronic properties of the pristine monolayer and layered bulk MoS$_2$ against other computational and experimental studies. In Sec.~\ref{formation-energy}, we present the defect formation energies and CTLs for the S vacancy defect in monolayer and layered bulk MoS$_2$ and demonstrate the effectiveness of the Freysoldt-Neugebauer 2D charge correction scheme when applied to these systems. We predict that the S vacancy in both systems is most stable in either the neutral or $-1$ charged states, in agreement with other studies in literature, validating our approach. In Sec.~\ref{electronic-struct}, we analyze the electronic structures of the S vacancy in the various charge states. The $-1$ charged S vacancy is found to undergo a Jahn-Teller distortion, which stabilizes this defect in both monolayer and layered bulk MoS$_2$.

\section{Computational Details} \label{comp-details}

We compute the material and defect properties using density functional theory (DFT) with the projector-augmented wave (PAW) method \cite{PAW-blochl1994, PAW-kresse1999} as implemented in the plane-wave code VASP \cite{vasp-kresse1996}. The PAW potentials describe the core states of Mo and S by the electronic configurations of [Ar]$4s^23d^{10}$ and [Ne], respectively. We treat the exchange-correlation using two different sets of functionals -- the Perdew-Burke-Ernzerhof (PBE) \cite{PBE-perdew1996} generalized gradient approximation (GGA) functional and the strongly constrained and appropriately normed (SCAN) \cite{SCAN-sun2015} meta-GGA functional -- and compare the results. For the calculations with SCAN, we also include long-range van der Waals interactions via the SCAN+rVV10 functional \cite{SCANrVV10-peng2016}. We perform spin-polarized calculations employing a plane-wave cutoff energy of 520~eV, which ensures energy convergence to within 1 meV/atom. To facilitate rapid convergence of the Brillion zone integration, we use Methfessel-Paxton smearing \cite{methfessel-paxton1989} with a smearing energy width of 0.10~eV and a $\Gamma$-centered Monkhorst-Pack $k$-point meshes \cite{monkhorst-pack1976}. For the structural relaxations, we use $k$-point meshes corresponding to a $12\times12\times1$ $k$-point mesh per hexagonal unit cell for monolayer MoS$_2$ ($\approx$ 400 $k$-points per reciprocal atom in 2D) and $12\times12\times3$ $k$-point mesh per hexagonal unit cell for layered bulk MoS$_2$ ($\approx$ 2500 $k$-points per reciprocal atom in 3D). For the density of states calculations, we double the density of the $k$-point meshes in all directions and use Gaussian smearing with a reduced smearing width of 0.02~eV.

We model the defective systems by constructing $3\times3\times1$, $4\times4\times1$, and $5\times5\times1$ supercells based on the hexagonal primitive unit cell, as well as $3\times2\times1$ and $4\times2\times1$ supercells based on an orthorhombic unit cell, and removing a single S atom to create a S vacancy. For monolayer MoS$_2$, in addition to varying the in-plane supercell size, we also vary the amount of vacuum spacing between layers to be 10, 15, or 20~\AA. Spin-orbit coupling was considered in a few select calculations and was found to change the defect formation energies by only 10 to 15~meV and to lead to a splitting of about 140~meV at the top of the valence band and of about 70~meV of the S vacancy defect level in monolayer MoS$_2$. While not insignificant, the effects are not so large as to qualitatively change our conclusions; therefore, the following results presented in this paper are reported without including spin-orbit coupling.

The formation energy $E^f[X^q]$ of a point defect $X$ with charge $q$ is determined from DFT calculations using a supercell approach following 
\begin{equation} \label{Eform}
E^f[X^q] = E_{\rm{tot}}[X^q] - E_{\rm{tot}}[\textrm{pristine}] - \sum_i n_i \mu_i + qE_{\textrm{F}} + E_{\textrm{corr}},
\end{equation} 
where $E_{\textrm{tot}}[X^q]$ and $E_{\textrm{tot}}[\textrm{pristine}]$ are the total DFT-derived energies of the supercell containing the defect $X$ and the pristine supercell respectively, $n_i$ is the number of atoms of species $i$ added/removed by the defect, $\mu_i$ is the corresponding chemical potential of the species, and $E_{\textrm{F}}$ is the Fermi energy. In this work, we considered only the Mo-rich/S-poor limit, for which the appropriate S chemical potential $\mu_\textrm{S}(\text{S-poor}) = (\mu_{\textrm{MoS}_2} - \mu_\textrm{Mo(bcc)})/2$. The final term in Eq.~\eqref{Eform}, $E_{\textrm{corr}}$, contains corrections to the formation energy due to electrostatic interactions with periodic images and compensating background charges, which are introduced in supercell calculations using plane-wave DFT approaches. Various correction schemes have been developed for charged defects in bulk 3D materials \cite{leslie1985,makov1995,lany2008,lany2009,freysoldt2009,freysoldt2011,komsa2012prb,kumagai2014,wu2017}. In this work, we use the approach developed by Freysoldt, Neugebauer, and Van de Walle \cite{freysoldt2009,freysoldt2011} to study the S vacancy in layered bulk MoS$_2$. 

Charged defects in single-layer materials pose additional challenges that lead to the divergence of the energy with vacuum spacing. Komsa \textit{et al.} proposed a uniform scaling scheme for charged defects at surfaces and interfaces \cite{komsa2013} and in 2D materials \cite{komsa2014}. In this work, we study the S vacancy in monolayer MoS$_2$ using the correction scheme developed recently by Freysoldt and Neugebauer \cite{freysoldt2018}.
The Freysoldt-Neugebauer scheme uses a surrogate model to directly correct the electrostatic energy induced by the wrong electrostatic boundary conditions. This scheme is computationally efficient as it is implemented as a post-processing step, requiring as input only the electrostatic potential of the converged DFT calculations. The correction scheme proposed by Freysoldt and Neugebauer has the advantage of not requiring any extrapolation or making assumptions about the finite-size scaling behavior, which would require the evaluation of large supercells to recover the correct asymptotic behavior~\cite{komsa2018-erratum}. 

\begingroup\squeezetable
\begin{table*}[htb]
\caption{Lattice constants, band gaps, and dielectric coefficients for monolayer and layered bulk MoS$_2$ calculated using different functionals, compared with experimental values. The PBE and SCAN+rVV10 values are calculated in this work, while the HSE values are taken from the literature. The in-plane lattice parameter $a$ is well-reproduced by the PBE and SCAN functionals, with SCAN+rVV10 also reproducing the $c$ lattice parameter (i.e., the interlayer spacing) in the layered bulk MoS$_2$. PBE and HSE functionals fail to reproduce this interlayer distance due to missing van der Waals interactions, therefore the $c$ lattice parameter is fixed to the experimental value (values marked with *). As expected, both PBE and SCAN functionals significantly underestimate the band gaps by about 0.8 to 1.0 eV in the monolayer and 0.3 to 0.4 eV in the layered bulk.}
\begin{ruledtabular}
\begin{tabular}{l>{\hspace{10pt}}l>{\hspace{10pt}}l>{\hspace{10pt}}l>{\hspace{10pt}}l>{\hspace{10pt}}l>{\hspace{10pt}}l}
 & \textit{a} (\AA) & \textit{c} (\AA) & $d_{\rm{S-S}}$ (\AA)& $E_{\rm{gap}}$ (eV) & $\varepsilon_{\parallel}$ $(\varepsilon_0)$ & $\varepsilon_{\bot}$ $(\varepsilon_0)$ \\
\hline
ML MoS$_2$: &  &  &  &  &  \\
\hspace{10pt}PBE & 3.18 & -- & 3.12 & 1.67 & \multicolumn{2}{c}{17.18} \\
\hspace{10pt}SCAN+rVV10 & 3.17 & -- & 3.11 & 1.80 & \multicolumn{2}{c}{16.27}\\
\hspace{10pt}HSE & 3.16\cite{komsa2015} & -- & & \makecell[l]{2.17, 2.21\cite{komsa2015}, \\ 2.25\cite{ding2011}, 2.3\cite{ellis2011}}
&  &  \\
\hspace{10pt}Expt. & 3.2$\pm$0.1\cite{huang2015,vancso2016} & -- & & \makecell[l]{2.40$\pm$0.05\cite{huang2015}, \\ 2.63\cite{soklaski2014}, 2.7\cite{krane2016}} &  &  \\
\hline
bulk MoS$_2$: &  &  &  &  &  \\
\hspace{10pt}PBE@expt. \textit{c} & 3.18 & 12.30* & 3.12 & 0.88 & 15.39 & 6.51 \\
\hspace{10pt}SCAN+rVV10 & 3.16 & 12.30 & 3.11 & 0.98 & 14.80 & 5.15 \\
\hspace{10pt}HSE@expt. \textit{c} & 3.16\cite{komsa2015} & 12.30*\cite{komsa2015} & & 1.45, 1.47\cite{komsa2015}, 1.50 \cite{ellis2011} &  &  \\
\hspace{10pt}Expt. & 3.15\cite{wakabayashi1975}, 3.16\cite{boker2001} & 12.29\cite{boker2001}, 12.3\cite{wakabayashi1975} & & \makecell[l]{1.17\cite{kautek1980}, 1.2\cite{goldberg1975,fortin1982}, \\ 1.23\cite{kam1982}, 1.29\cite{gmelin1995}} 
& 15.2$\pm$0.2\cite{wieting1971} & 6.2$\pm$0.1\cite{wieting1971} \\
\end{tabular}
\end{ruledtabular}
\label{tab:matprops}
\end{table*}
\endgroup

Table~\ref{tab:matprops} compares the lattice parameters, band gaps, and dielectric coefficients for monolayer and layered bulk MoS$_2$ computed in this work with experimental values. The PBE and SCAN+rVV10 values are calculated in this work, while the HSE values are taken from Refs.~\onlinecite{komsa2015,ding2011,ellis2011}. All functionals well reproduce the in-plane lattice parameter $a$, with SCAN+rVV10 also reproducing the experimental $c$ lattice parameter (i.e., the interlayer spacing) in the layered bulk MoS$_2$. PBE and HSE functionals significantly overpredict the interlayer distance in layered bulk MoS$_2$ due to missing van der Waals interactions; therefore the $c$ lattice parameter is fixed to the experimental value in subsequent calculations. As expected, both PBE and SCAN functionals significantly underestimate the fundamental band gaps by about 30--40\% in the monolayer \cite{huang2015,soklaski2014,krane2016} and 20--30\% in the layered bulk \cite{kautek1980,goldberg1975,fortin1982,kam1982,gmelin1995}. Some studies report good agreement between PBE-computed band gaps and experimentally-measured \textit{optical} band gaps; however, this is misleading as the appropriate comparison is with the \textit{fundamental} band gap as is reported here. Despite underestimating the fundamental band gaps, it is still worth noting that PBE and SCAN do qualitatively reproduce the key features of the band structure, showing the indirect-to-direct band gap transition when going from the layered bulk to monolayer MoS$_2$.

The charge correction scheme requires the dielectric properties as input~\cite{freysoldt2009, freysoldt2011, freysoldt2018}. We compute the dielectric tensor components for the monolayer with DFT using supercells containing a slab of thickness $d^{\rm{slab}}$ and a vacuum region of thickness $d^{\rm{vac}}$ ($d^{\rm{slab}}$ + $d^{\rm{vac}}$ = $d^{\rm{sc}}$ = supercell $c$ lattice parameter). As a result, the computed dielectric tensor components for the supercell $\varepsilon^{\rm{sc}}$ are combinations of the dielectric tensor components for the monolayer $\varepsilon^{\rm{slab}}$ and vacuum $\varepsilon^{\rm{vac}}$ (= 1). The in-plane components (subscripted $\parallel$) behave as capacitors in parallel, while the out-of-plane components (subscripted $\bot$) behave as capacitors in series, yielding the following relations~\cite{freysoldt2008}:
\begin{equation}
\varepsilon_{\parallel}^{\rm{sc}} 
= \frac{d^{\rm{vac}}}{d^{\rm{sc}}} \varepsilon^{\rm{vac}} + \frac{d^{\rm{slab}}}{d^{\rm{sc}}} \varepsilon_{\parallel}^{\rm{slab}}
= 1 + \left (\varepsilon_{\parallel}^{\rm{slab}} - 1 \right )\frac{d^{\rm{slab}}}{d^{\rm{sc}}}
\end{equation}
\begin{equation}
\frac{1}{\varepsilon_{\bot}^{\rm{sc}}}
= \frac{d^{\rm{vac}}}{d^{\rm{sc}}} \frac{1}{\varepsilon^{\rm{vac}}} + \frac{d^{\rm{slab}}}{d^{\rm{sc}}} \frac{1}{\varepsilon_{\bot}^{\rm{slab}}}
= 1 + \left (\frac{1}{\varepsilon_{\bot}^{\rm{slab}}} - 1 \right ) \frac{d^{\rm{slab}}}{d^{\rm{sc}}}.
\end{equation}
The dielectric tensor components for the slab $\varepsilon_{\parallel}^{\rm{slab}}$ and $\varepsilon_{\bot}^{\rm{slab}}$ are only uniquely defined for a given choice of slab thickness $d^{\rm{slab}}$. To solve for a unique combination of slab thickness and dielectric constants, we make the simplifying assumption that $\varepsilon_{\parallel}^{\rm{slab}} = \varepsilon_{\bot}^{\rm{slab}}$, i.e., that the slab is dielectrically isotropic. This gives the following expression for the slab dielectric constant~\cite{freysoldt2008},
\begin{equation}
\varepsilon^{\rm{slab}} = \frac{\varepsilon_{\parallel}^{\rm{sc}} - 1}{1 - (\varepsilon_{\bot}^{\rm{sc}})^{-1}}
\end{equation}
and the corresponding slab thickness,
\begin{equation}
d^{\rm{slab}} = d^{\rm{sc}} \, \left (\frac{1}{1 - (\varepsilon_{\bot}^{\rm{sc}})^{-1}} + \frac{1}{1 - \varepsilon_{\parallel}^{\rm{sc}}}\right )^{-1}.
\end{equation}
Based on these expressions, we estimate the averaged dielectric constant and slab thickness to be 17.2 $\varepsilon_0$ and 5.4~\AA\ computed with PBE and 16.3 $\varepsilon_0$ and 5.4~\AA\ computed with SCAN+rVV10. These values of slab thickness are physically meaningful as they correspond approximately to the S--S distance ($d_{\mathrm{S-S}}$ listed in Table~\ref{tab:matprops}) + $2 \times$ the covalent radius of S ($r_{\mathrm{S}} = 1.05$ \AA).

The assumption of an isotropic dielectric slab need not be valid for a monolayer; however, this simplified dielectric model correctly reproduces the asymptotic screening properties of the repeated slab system. The consistent choice of $d^{\rm{slab}}$ and $\varepsilon^{\rm{slab}}$ ensures that, despite the approximation of mapping the anisotropic dielectric tensor of the slab onto an isotropic dielectric tensor, the correct asymptotic behavior of the electrostatic potential is still recovered. At distances shorter or comparable to the length scale set by $d^{\rm{slab}}$, both the dielectric anisotropy within the layer as well as microscopic variations due to the detailed atomic and electronic structure come into play. Since the latter is unavoidable and fundamentally limits all continuum modeling attempts, we do not expect that a more elaborate setup of the surrogate model would significantly improve its performance.

Calculating the dielectric tensor for the layered bulk MoS$_2$ is much more straightforward since there is no vacuum region in the cells. The dielectric tensors are evaluated with the PBE functional using density functional perturbation theory (DFPT), and with the SCAN+rVV10 functional using finite field method as DFPT is not currently implemented in \textsc{vasp} for meta-GGAs and hybrid functionals. The in-plane and out-of-plane static dielectric coefficients for layered bulk MoS$_2$ computed in this work agree well with experimentally-measured values \cite{wieting1971}. Applying the simple dielectric model from above to the layered bulk system ($\varepsilon_{\parallel}^{\rm{slab}} = \varepsilon_{\bot}^{\rm{slab}} = \varepsilon^{\rm{slab}}$, $d^{\rm{slab, bulk}}$ = 2$d^{\rm{slab, ML}}$, $d^{\rm{sc}}$ = 12.3 \AA) predicts in-plane dielectric constants of 15.2 $\varepsilon_0$ (PBE) and 14.3 $\varepsilon_0$ (SCAN+rVV10), and out-of-plane dielectric constants of 5.8 $\varepsilon_0$ (PBE) and 5.5 $\varepsilon_0$ (SCAN+rVV10), within $\approx 10\%$ of the directly computed values, which further validates our use of the simple dielectric model for the monolayer.

The dielectric properties of each system are required as inputs to the charge correction schemes employed in this work. For the bulk charge correction scheme, we provide the full anisotropic dielectric tensor computed for layered bulk MoS$_2$. Meanwhile, the 2D charge correction scheme, as currently implemented, takes as inputs only a single averaged dielectric constant for the slab as well as a slab thickness, which defines the positions of the dielectric interfaces. We test the sensitivity of the charge correction scheme to the choice of dielectric constant -- changing the dielectric constant by up to 50\% changes the formation energies and CTLs by around 100 meV which does not qualitatively alter our results. 

\section{Defect formation energy} \label{formation-energy}

\begin{figure}[tb]
\centering
\subfloat{\includegraphics[width=3.0in]{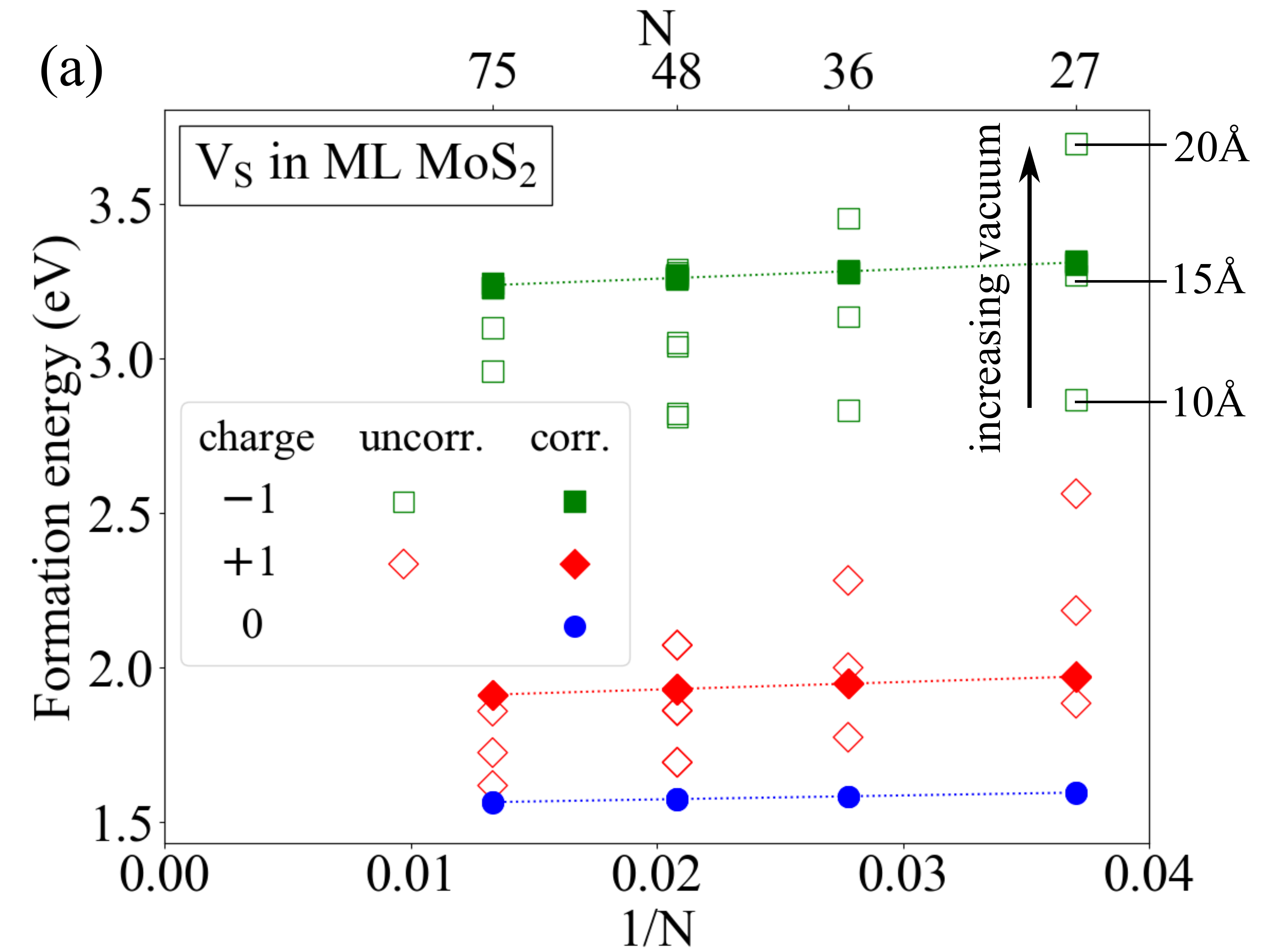}} \\
\subfloat{\includegraphics[width=3.0in]{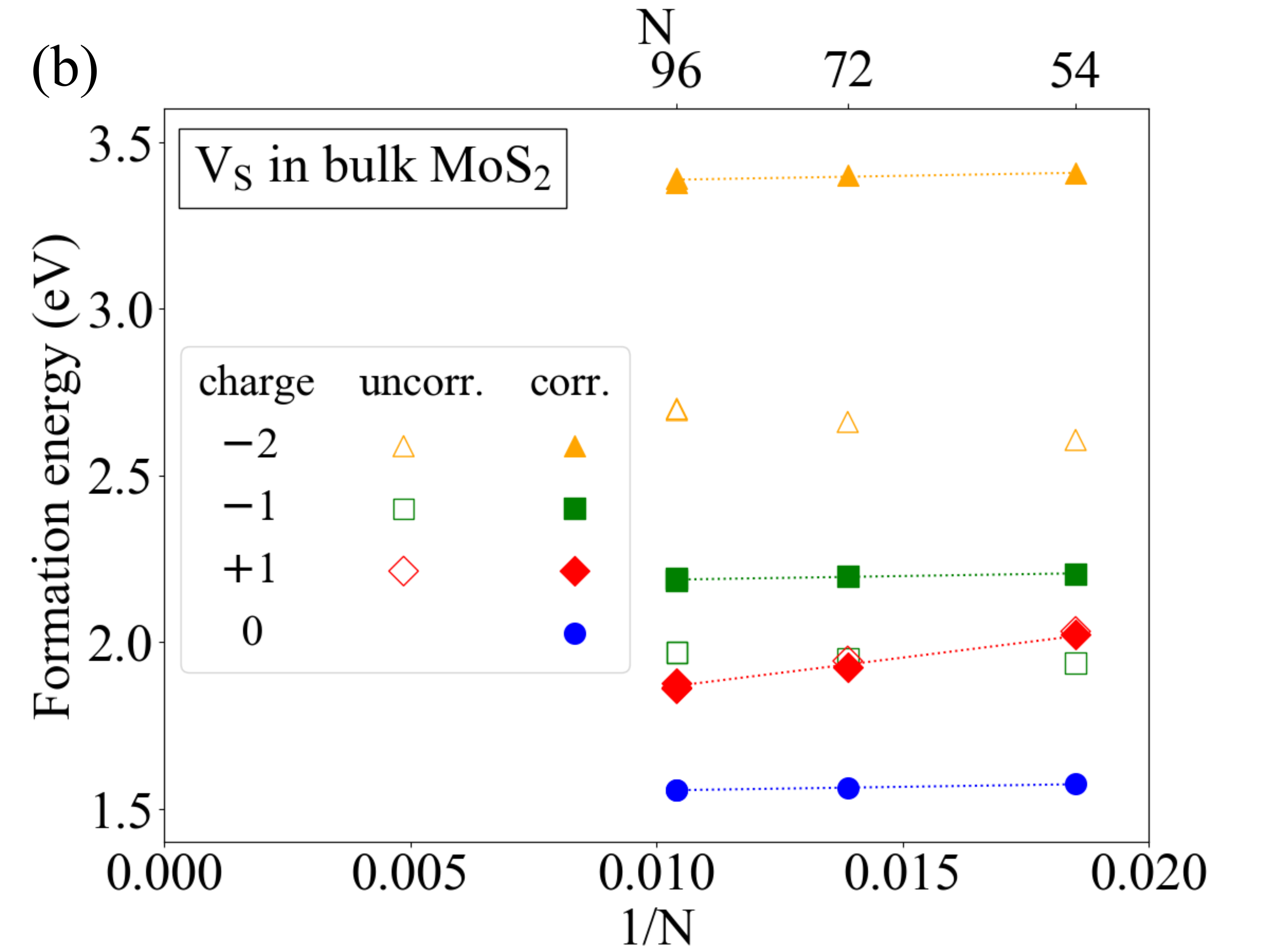}}
\caption{Uncorrected (open symbols) and corrected (filled symbols) formation energies for S vacancies in (a) monolayer MoS$_2$ and (b) layered bulk MoS$_2$ in different charge states, calculated with SCAN+rVV10. The Fermi level is set to the valence band maximum. In the top plot for S vacancies in monolayer MoS$_2$, the multiple data points corresponding to each supercell size indicate the energies evaluated in supercells with different vacuum spacings of 10, 15, and 20 \AA. The uncorrected energies diverge with increasing vacuum spacing and also exhibit a strong dependence on in-plane supercell size. The corrected energies are well converged across all supercell and vacuum sizes in all cases except the +1 S vacancy in layered bulk MoS$_2$, for which the correction does not work due to the delocalized nature of the defect state (see text).}
\label{fig:convergence}
\end{figure}

Figure~\ref{fig:convergence} demonstrates that upon application of the charge correction scheme, the defect formation energies for charged S vacancies in the monolayer and the layered bulk MoS$_2$ become well-converged in all cases except for the +1 S vacancy in layered bulk MoS$_2$. On both plots, the open symbols indicate the uncorrected defect formation energies in the Mo-rich/S-poor limit, calculated with SCAN+rVV10. The uncorrected energies strongly depend on the in-plane supercell size, and for the monolayers, also on the vacuum spacing. After correction, the energies are converged to within 100 meV across all supercell and vacuum sizes in all cases except the +1 S vacancy in layered bulk MoS$_2$, for which the correction does not work because the defect charge turns out to be delocalized (c.f. Fig.~\ref{fig:chgden_bulk_delocalized}). For delocalized defect states, the energy correction evaluated by the electrostatic correction scheme is not meaningful, and in this system turns out to be close to zero, leading to an overlap between the uncorrected and ``corrected" (red diamond) symbols in Fig.~\ref{fig:convergence}(b). The dotted lines connecting the corrected energies are included as a guide to the eye, showing that when the correction scheme is applied successfully, only a small dependence on in-plane supercell size, which is comparable to that for the neutral defect remains. This small supercell size dependence reflects the elastic interactions between defects, which are not accounted for in the electrostatic charge correction scheme. The unphysical linear divergence in energy with vacuum spacing is effectively addressed by the correction scheme as evidenced by the overlapping filled symbols in the Fig.~\ref{fig:convergence}(a). These plots also show that the magnitudes of the correction terms range from tens to hundreds of meV depending on the system, charge state, and supercell size, and must be included when evaluating formation energies of charged point defects.

\begin{figure}[htb]
\centering
\subfloat{\includegraphics[width=3.0in]{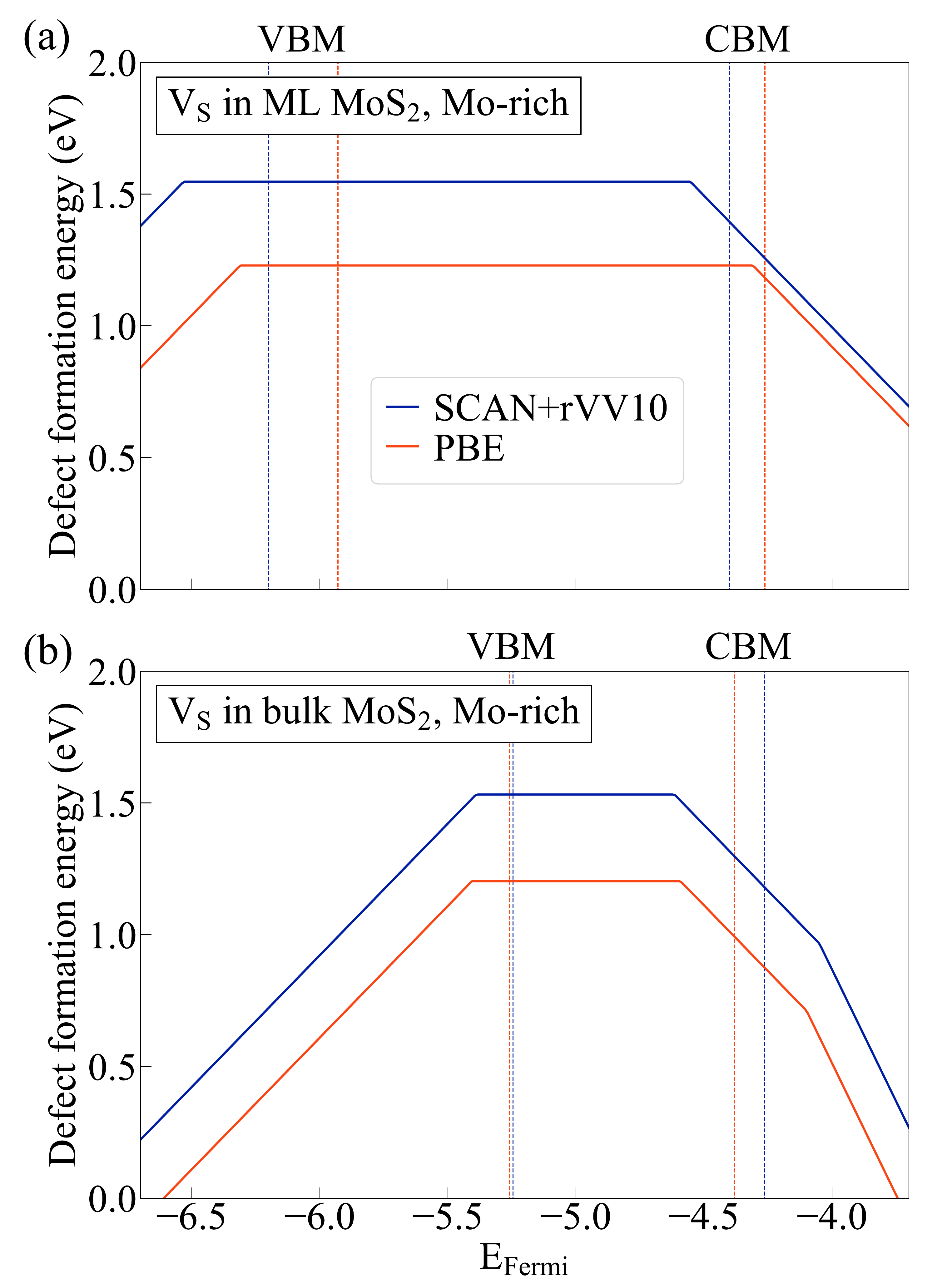}}
\caption{Formation energy of the S vacancy in (a) monolayer MoS$_2$ and (b) layered bulk MoS$_2$ as a function of Fermi level position, calculated with PBE (orange) and SCAN+rVV10 (blue). The valence and conduction band edge positions calculated with respect to vacuum level are indicated by the vertical dashed lines. The slope of the formation energy plot corresponds to the most stable charge state for the defect over that range of Fermi energies. In all cases, the neutral S vacancy is predicted to be most stable across most of the gap, and only the 0/$-1$ charge transition level is predicted to be within the band gap, close to the conduction band minimum.}
\label{fig:CTL}
\end{figure}

Figure~\ref{fig:CTL} shows that the S vacancy is most stable in the neutral charge state in both monolayer and layered bulk MoS$_2$ for Fermi energies spanning most of the band gap, with the $-1$ charged S vacancy becoming more favorable close to the conduction band minimum (CBM). The formation energy of the neutral S vacancy in both monolayer and layered bulk MoS$_2$ under Mo-rich/S-poor conditions is computed with SCAN+rVV10 to be $\approx$ 1.5 eV, in good agreement with other DFT-computed values reported in literature \cite{komsa2015,noh2014,naik2018}. PBE predicts the formation energy to be $\approx$ 0.3 eV lower; this is in part due to the underestimation of $\mu_{\textrm{MoS}_2}$ and therefore $\mu_\textrm{S}$ resulting from missing van der Waals interactions.

The kinks in the formation energy plots indicate charge transition levels (CTLs). The 0/$-1$ CTL is predicted to be within the band gap close to the CBM, corresponding to a deep acceptor state. Our calculations estimate the $0/-1$ CTL to be 1.6 to 1.7~eV above the valence band minimum (VBM) in monolayer MoS$_2$ and 0.6 to 0.7~eV above the VBM in layered bulk MoS$_2$, again in good agreement with previous studies \cite{komsa2015,noh2014,naik2018}. As previously noted, both PBE and SCAN functionals significantly underestimate the band gaps; therefore, some CTLs which appear outside the band gap in our calculations may fall within the band gap, leading to multiple defect levels within the gap. Indeed, calculations with HSE and GW in the literature have predicted that the $+1/0$ CTL in monolayer MoS$_2$ and the $-1/-2$ CTL in layered bulk MoS$_2$ may also fall within the band gap \cite{komsa2015,naik2018}. The position of the $+1/0$ CTL in layered bulk MoS$_2$ is uncertain since the correction method does not work when the extra charge occupies a delocalized state. Figure~\ref{fig:chgden_bulk_delocalized} shows that the extra hole in the calculation of the +1 charged S vacancy in layered bulk MoS$_2$ has a delocalized charge distribution that corresponds to an empty state at the VBM. This indicates that the localized defect state associated with the S vacancy is located within the valence band region (see Fig.~\ref{fig:pdos_bulk} in Sec.~\ref{electronic-struct}) and not likely to be stabilized within the band gap, hence the $+1/0$ CTL is not relevant. 

\section{Electronic structure} \label{electronic-struct}

\begin{figure*}[htb]
\centering
\subfloat{\includegraphics[width=6.5in]{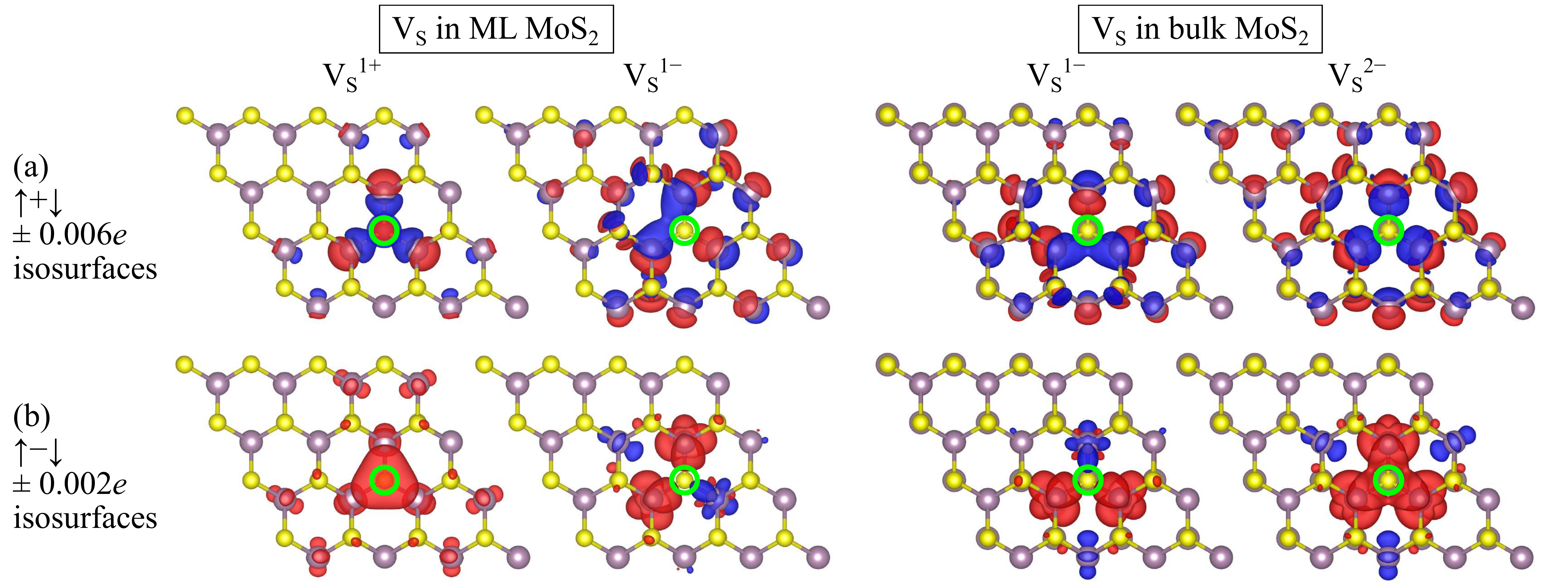}}
\caption{(a) Total charge and (b) spin density distributions of the extra hole or electron(s) around the charged S vacancy in monolayer and layered bulk MoS$_2$. Mo atoms are depicted in purple and S in yellow, the position of the S vacancy within each $4 \times 4$ supercell is marked by the green circles, and the red (blue) indicate the positive (negative) charge and spin isosurfaces. The symmetry is broken in the $-1$ charge state in both the monolayer and layered bulk systems.}
\label{fig:chgden}
\end{figure*}

\begin{figure*}[htb]
\centering
\subfloat{\includegraphics[width=6.0in]{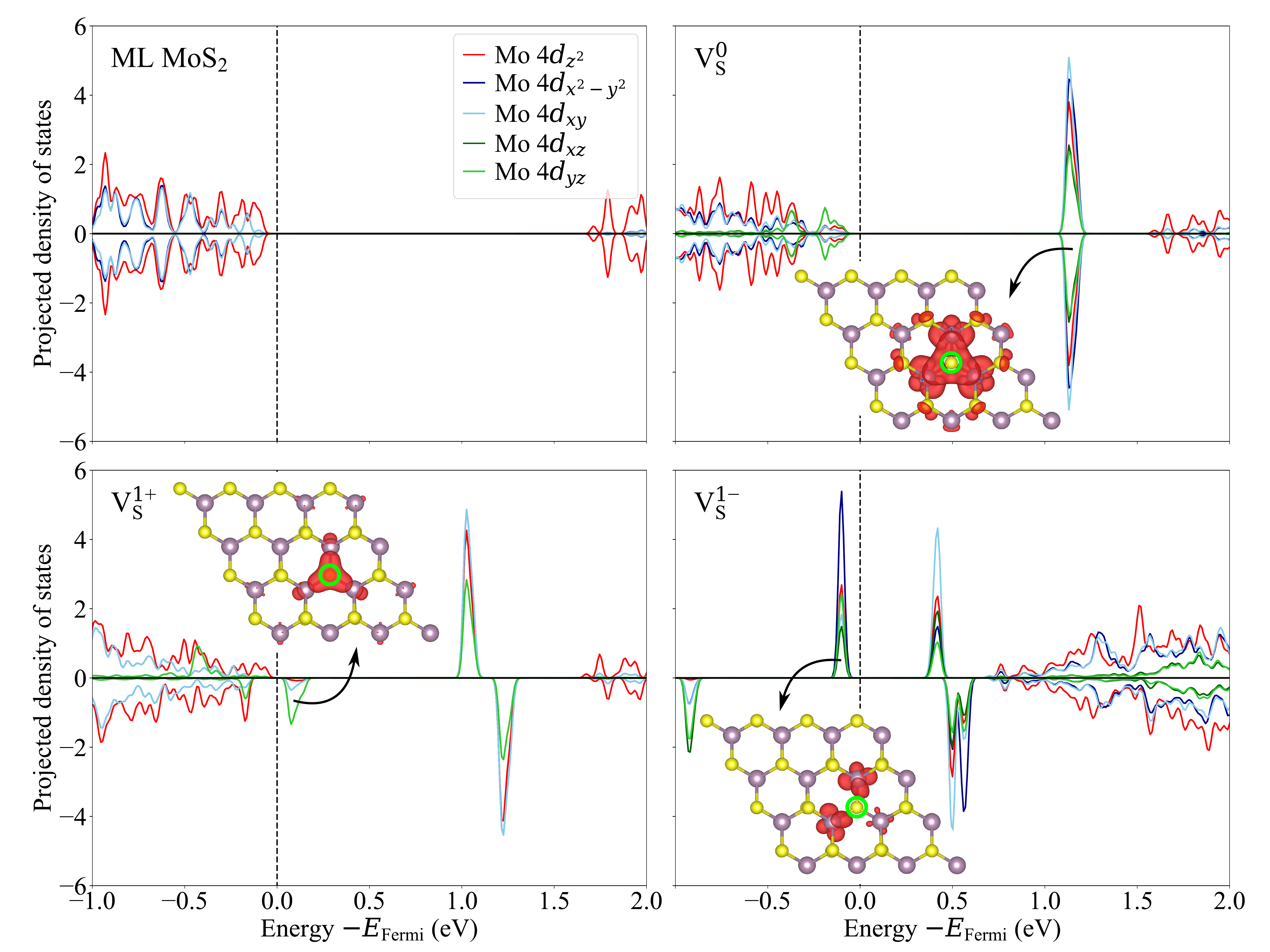}}
\caption{Projected density of states for the pristine monolayer MoS$_2$ and the S vacancy in monolayer MoS$_2$ in the neutral (V$_\textrm{S}^0$), $+1$ (V$_\textrm{S}^{1+}$), and $-1$ (V$_\textrm{S}^{1-}$) charge states. The density of states is projected onto the $d$-orbitals of the three Mo atoms directly adjacent to the S vacancy. The defect state in the band gap has primarily $d_{x^2-y^2}$ (dark blue) and $d_{xy}$ (light blue) character. In the $-1$ charge state, the degeneracy between these two orbitals is broken, with the additional electron occupying a state with dominant $d_{x^2-y^2}$ character. The insets illustrate the electronic orbitals corresponding to the defect states of interest.}
\label{fig:pdos_ML}
\end{figure*}

\begin{figure*}[htb]
\centering
\subfloat{\includegraphics[width=6.0in]{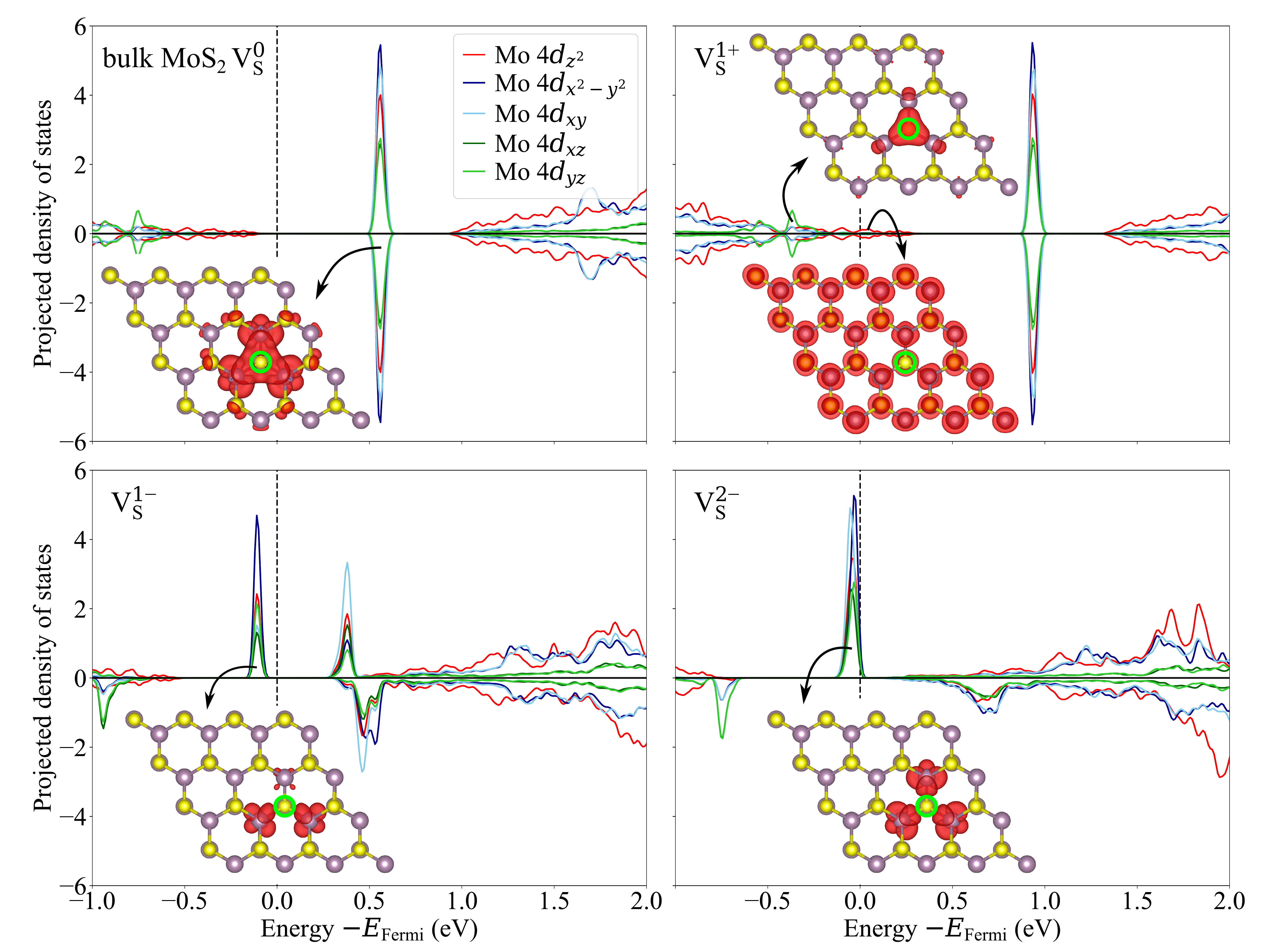}}
\caption{Projected density of states for the S vacancy in layered bulk MoS$_2$ in the neutral (V$_\textrm{S}^0$), $+1$ (V$_\textrm{S}^{1+}$), $-1$ (V$_\textrm{S}^{1-}$), and $-2$ (V$_\textrm{S}^{2-}$) charge states. The density of states is projected onto the $d$-orbitals of the three Mo atoms directly adjacent to the S vacancy. Similar to the S vacancy in the monolayer, the degeneracy between $d_{x^2-y^2}$ (dark blue) and $d_{xy}$ (light blue) orbitals is broken in the $-1$ charge state. The degeneracy is restored in the $-2$ charge state, for which the parallel spin configuration is lower in energy. The insets show the electronic orbitals corresponding to the defect states of interest.}
\label{fig:pdos_bulk}
\end{figure*}

Figure~\ref{fig:chgden} shows the total charge and spin density distributions around the charged S vacancy in monolayer and layered bulk MoS$_2$. The charge and spin densities associated with the additional hole or electron(s) remain fairly localized around the defect site for the $+1$ and $-1$ charged S vacancy in the monolayer, as well as for the $-1$ and $-2$ charged S vacancy in the bulk. The charge and spin densities around the $-1$ charged S vacancy look very similar in both the monolayer and layered bulk, demonstrating a breaking of the 3-fold symmetry of the native lattice. The 3-fold symmetry around the defect site is maintained in all the other charged and neutral S vacancy configurations we studied. Each of the configurations depicted in Fig.~\ref{fig:chgden} exhibits a net magnetic moment, including the $-2$ charged S vacancy in the bulk for which the parallel spin configuration is more stable than the anti-parallel (non-magnetic) configuration by $\approx$ 150 meV when evaluated with SCAN+rVV10 ($\approx$ 40 meV when evaluated with PBE). This is also reflected in the projected density of states plots in Figs. \ref{fig:pdos_ML} and \ref{fig:pdos_bulk}.

\begin{figure}[htb]
\centering
\subfloat{\includegraphics[width=3.0in]{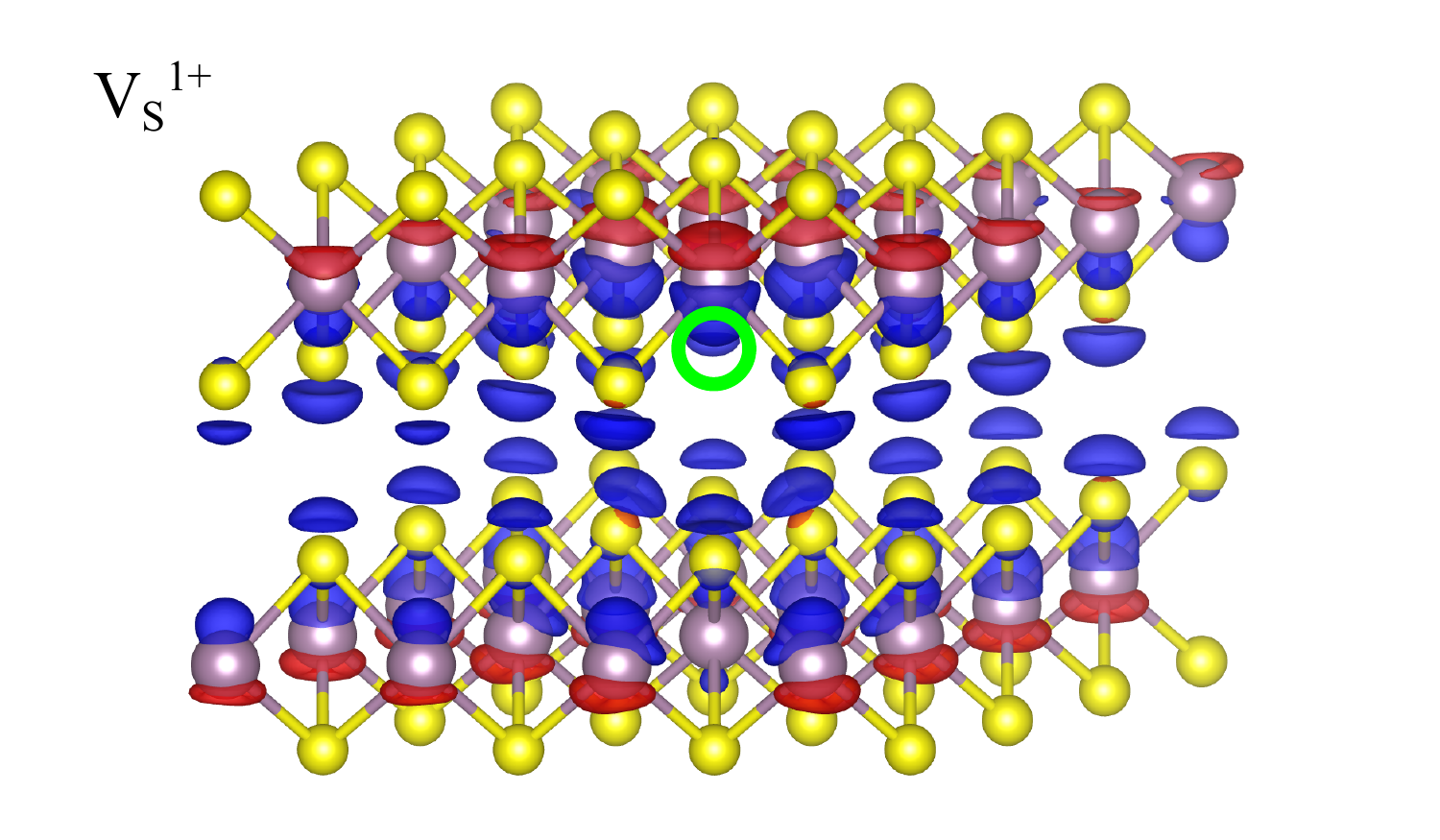}}
\caption{Side view of the total charge density distribution of the extra hole around the positively charged S vacancy in layered bulk MoS$_2$. Mo atoms are depicted in purple and S in yellow, the position of the S vacancy within the $4 \times 4$ supercell is marked by the green circle, and the red (blue) indicate the positive (negative) $0.001 e$ charge isosurfaces. The charge density is completely delocalized not only within the layer containing the S vacancy but also on adjacent layers.}
\label{fig:chgden_bulk_delocalized}
\end{figure}

Figure~\ref{fig:chgden_bulk_delocalized} illustrates the delocalization of charge in the $+1$ S vacancy in layered bulk MoS$_2$, explaining why the charge correction scheme -- which assumes a relatively localized charge -- does not work in this case. Unlike in monolayer MoS$_2$, in layered bulk MoS$_2$ the extra charge (hole) associated with the $+1$ S vacancy is completely delocalized, not only within the layer containing the S vacancy but also over adjacent layers. Since the correction scheme assumes a relatively localized charge, it fails in this case, as seen in Fig.~\ref{fig:convergence}. As a result, we are unable to accurately quantify the formation energy of this defect. However, the delocalized nature of the charge as well as the calculated density of states for this defect (Fig. \ref{fig:pdos_bulk}b), which indicates that a state at the top of the valence band is depleted when the $+1$ S vacancy is formed, are consistent with it being a state within the valence band.

The projected densities of states corresponding to the S vacancy in different charge states in monolayer MoS$_2$ (Fig.~\ref{fig:pdos_ML}) and layered bulk MoS$_2$ (Fig.~\ref{fig:pdos_bulk}) show that a defect state with primarily Mo $d_{x^2-y^2}$ and $d_{xy}$ orbital character is induced in the gap. The density of states is projected onto the $d$-orbitals of the three Mo atoms directly adjacent to the S vacancy. In defect-free MoS$_2$, the Mo atoms have trigonal prismatic ($D_{3\textrm{h}}$) symmetry, which gives rise to the following energetic splitting of $d$-orbitals: $d_{z^2}$ ($a_1'$ orbital) $< d_{x^2-y^2} = d_{xy}$ ($e'$ orbitals) $< d_{xz} = d_{yz}$ ($e''$ orbitals). The degeneracy of the orbitals is reflected in the overlapping $d$-orbital contributions in the projected density of states for the pristine monolayer for the blue $d_{x^2-y^2}$ and $d_{xy}$ states ($e'$) and green $d_{xz}$ and  $d_{yz}$ ($e''$) peaks. When a S vacancy is created, this generates a doubly degenerate defect state in the band gap with primarily $e'$ character. Another defect state with primarily $e''$ character also appears near the top of the valence band. In the neutral state (V$_\textrm{S}^0$), the state at the top of the valence band is filled while the states in the gap remain empty. 

When adding an extra electron to the neutral S vacancy, the system undergoes a Jahn-Teller distortion, which stabilizes the $-1$ charged S vacancy in both monolayer and layered bulk MoS$_2$. In the negatively charged state (V$_\textrm{S}^{1-}$), the previously doubly degenerate defect state in the gap splits and the additional electron occupies the lowest energy state, which exhibits predominantly $d_{x^2-y^2}$ (dark blue) character. This breaking of the degeneracy among the $e'$ orbitals is also associated with the breaking of the $D_{3\textrm{h}}$ symmetry of the vacancy. Such simultaneous electronic and geometric symmetry-breaking is an example of a Jahn-Teller distortion, and is responsible for stabilizing the $-1$ charged S vacancy in both monolayer and layered bulk MoS$_2$. The symmetric V$_\textrm{S}^{1-}$ structures without Jahn-Teller distortions are higher in energy by 116~meV (43~meV) and 120~meV (50~meV) in the monolayer and layered bulk MoS$_2$, respectively, when evaluated with the SCAN+rVV10 (PBE) functional.

Unlike in the neutral and negatively charged S vacancy, the defect state associated with the positively charged S vacancy (V$_\textrm{S}^{1+}$) appears to have a different character in the monolayer compared to in the layered bulk. In the monolayer, the electron is removed from a localized defect state near the top of the valence band. However, in the layered bulk, Fig.~\ref{fig:pdos_bulk} shows that the electron removal depletes a delocalized state at the top of the valence band rather than the localized defect state, which is located 0.6~eV below the VBM.

This difference in behavior between the monolayer and bulk for the V$_\textrm{S}^{1+}$ vacancy is notable when comparing the predicted +1/0 CTLs (see Fig.~\ref{fig:CTL}). Indeed, the +1/0 CTL is predicted to fall below the VBM extracted from the Kohn-Sham band structures in both cases. Therefore, in both cases, the delocalized valence-band-like hole should become more stable than charge localization on the defect under appropriate electrostatic boundary conditions. This suggests that the charge localization may be an artefact of the calculation due to the periodic boundary conditions which artificially stabilize the observed localized solution compared to the delocalized one.

To estimate the energy correction for a delocalized charge, we consider a surrogate model, similar to the Gaussian charge model, where the excess charge is delocalized in the plane. We find that the uncorrected energy of the delocalized solution does indeed appear to be higher than that of the localized solution. Therefore, DFT calculations for charged simulation cells can converge to a localized solution, even though after correction, the delocalized solution might be the actual ground state. This systematic error arises because the periodic energy of a localized charge with a compensating homogeneous background is always attractive, while the energy of a delocalized charge with the same background charge density approaches zero.

The same argument also explains the apparent discrepancy between the position of the predicted -1/-2 CTL in layered bulk MoS$_2$ and the location of the corresponding defect states in the density of states. When a second electron is added to create the $-2$ charged S vacancy (V$_\textrm{S}^{2-}$) in layered bulk MoS$_2$, the $D_{3\textrm{h}}$ symmetry is restored, and the most stable configuration is found to be when the two extra electrons have their spins aligned ($\mu_{\rm{B}} = 2$). The density of states for V$_\textrm{S}^{2-}$ in layered bulk MoS$_2$ -- depicted in the bottom right panel of Fig.~\ref{fig:pdos_bulk} -- suggests that the second additional electron also occupies a localized defect state within the band gap. However, based on the defect formation energies, we predicted the -1/-2 CTL in layered bulk MoS$_2$ to be above the CBM. Again, due to the compensating homogeneous background, the localized defect state corresponding to V$_\textrm{S}^{2-}$ is artificially stabilized and therefore shows up in the gap in the density of states, which is not corrected. When the correct electrostatic boundary conditions are accounted for by including the correction term in the defect formation energies, the energy of the defect state is raised by $\approx$ 0.7 to 0.8~eV (c.f. Fig.~\ref{fig:convergence}(b)), pushing it into the conduction band.

These implicit consequences of the homogeneous charge background imply that one must be careful when basing the analysis of defect states \textit{solely} on calculations of the electronic structure -- such as the density of states -- which is quite often the case in the literature. In Kohn-Sham DFT, band structure and density of states are calculated based on one-electron energies, which is an approximation of the actual interacting electron system. While these calculations can provide valuable insight into the defect states and their orbital contributions, they may not be the most accurate way to estimate the positions of these defect states and CTLs. The location of the CTLs is given by the energy difference between different charge states of the defects and, as such, includes exchange and correlation contributions that are beyond the single-particle energies of the density of states. This many-body contribution can shift the energy of the CTLs. Moreover, as we argue above, there are some cases for which DFT may predict the wrong electronic ground state when a compensating homogeneous background charge is included, which can lead to incorrect conclusions being drawn based on the density of states alone. Furthermore,
different choices of functionals may also shift the CTLs and band edge positions relative to the vacuum level by up to 0.5~eV or more, which also adds to the uncertainty in predicting the energy of the CTLs relative to the band edges. \textit{A posteriori} correction schemes for band alignment \cite{Alkauskas2011,freysoldt2016} may be used in combination with the current charge correction scheme to obtain more accurate estimates of the positions of the defect states. Therefore, it is essential to employ and compare different approaches for investigating defect properties –- as presented in this work -– to obtain a more reliable and complete understanding of these defects.

Finally, we compare our electronic structure analysis with previous studies. Based on our electronic structure calculations, we identify the defect state near the top of the valence band to have a predominantly $e''$ character. The corresponding electronic orbital depicted in the inset of the V$_\textrm{S}^{1+}$ subplot of Fig. \ref{fig:pdos_ML} is largely localized in the region between the three Mo atoms neighboring the S vacancy site, which does suggest significant contributions from the $d_{xz}$ and $d_{yz}$ orbitals with lobes oriented in those directions. Our findings are in contrast to previous studies~\cite{noh2014,lu2018}, which identified that state as the singlet $a_1'$ (i.e., $d_{z^2}$) state instead. These same studies did identify the defect state in the gap to be the doubly degenerate $e'$ state, in agreement with our results. Noh et al. \cite{noh2014} also computed the density of states for V$_\textrm{S}^{1-}$, and while they did find a shift between the up and down-spin states as we did, they did not observe the splitting of the defect state indicative of a Jahn-Teller distortion. Komsa and Krasheninnikov \cite{komsa2015} briefly mention that the system undergoes a Jahn-Teller distortion upon addition of an extra electron; however, they did not elaborate further on that assertion or present any analysis of the defect electronic structure.

\section{Conclusion} \label{conc}

In this work, we performed density functional theory calculations to compute the formation energies and charge transition levels associated with sulfur vacancies in monolayer and layered bulk MoS$_2$.
We utilize the correction scheme recently developed by Freysoldt and Neugebauer to ensure the appropriate electrostatic boundary conditions for charged defects in 2D materials.
We demonstrate the effectiveness of the correction scheme for the convergence of the defect formation energies with respect to vacuum spacing and in-plane supercell dimensions. 
We benchmark the SCAN+rVV10 functional and this new charge correction scheme for 2D monolayers against other studies in literature, and find good agreement, validating our approach.
By analyzing the electronic structures of the defects in different charge states, we predict that both systems undergo a Jahn-Teller distortion, which helps stabilize the sulfur vacancy in the $-1$ charged state.

We show that the ubiquitous finite-size errors in charged-defect calculations for 2D materials tend to favor localized solutions over delocalized ones, irrespective of the sign of the correction applied to the localized state, and independent of the chosen \textit{a posteriori} correction scheme. As a result, the uncorrected stability region may fall within the band gap, allowing us to apply the charge correction scheme to estimate the CTLs even if they fall (after correction) outside the band gap. The results illustrate that the combination of the \textit{a posteriori} charge correction schemes with computationally feasible functionals provides a valuable tool for predicting the properties of charged defects in 2D semiconductor materials.



\begin{acknowledgments}
This work was supported by the National Science Foundation under the awards DMR-1748464, DMR-1539916, and OAC-1740251.
Computational resources were provided by the University of Florida Research Computing Center.
\end{acknowledgments}


%

\end{document}